\documentclass[12pt,twoside,fleqn]{article}
\usepackage{epsfig}
\usepackage{psfig}
\def\lsim{\raise0.3ex\hbox{$<$\kern-0.75em\raise-1.1ex\hbox{$\sim$}}}
\def\gsim{\raise0.3ex\hbox{$>$\kern-0.75em\raise-1.1ex\hbox{$\sim$}}}
\makeatletter
\@addtoreset{equation}{section}
\makeatother

\newcommand{\beqn} {\begin{equation}}
\newcommand{\eqn} {\end{equation}}

\newcommand{\slsh}[1] {#1\kern-.43em/}
\newcommand{\real}{{\sf I}\kern-.12em{\sf R}}
\newcommand{\comp}{{\sf I}\kern-.48em{\sf C}}
\newcommand{\nin} {\in\kern-.6em/}
\newcommand{\Tr} {\mbox{Tr}}

\def\MEF{m_{\rm eff}}\def\mef{\ifmmode\MEF\else$\MEF$\fi}

\begin{document}
\thispagestyle{empty}
%
 \mbox{} \hfill BI-TP 96/19\\
 \mbox{} \hfill May 1996\\
\begin{center}
\vspace*{1.0cm}
{{\large \bf The ${\cal O}(g^6)$ coefficient in the thermodynamic
potential of hot} \\
\medskip
{\large \bf SU(N) Gauge Theories and MQCD}
 } \\
\vspace*{1.0cm}
{\large F. Karsch$^1$, M. L\"utgemeier$^1$, A. Patk\'os$^2$
and J. Rank$^1$}
\\
\vspace*{1.0cm}
{\normalsize
$\mbox{}$ {$^1$ Fakult\"at f\"ur Physik, Universit\"at Bielefeld, P.O. Box 100131,
D-33501 Bielefeld, Germany  \\
$^2$ Institute of Physics, E\"otv\"os University, Budapest, Hungary
}}\\
\vspace*{1.8cm}
{\large \bf Abstract}
\end{center}

The non-perturbative input necessary for the determination of 
the ${\cal O}(g^6)$
part of the weak coupling expansion of the free energy density for SU(2) and
SU(3) gauge theories is estimated. Although the perturbative information
completing the contribution to this order is missing, we give 
arguments that the magnetic fluctuations are dominated by
screened elementary magnetic gluons.
\setlength{\baselineskip}{1.3\baselineskip}

\newpage
\setcounter{page}{1}
\section{Introduction}

Perturbation theory in its original formulation fails beyond ${\cal O}(g^5)$
in the calculation of the free energy density of non-Abelian gauge theories
\cite{Lin80}. Very recently Braaten \cite{Bra95} has pointed out that the
coefficients of the higher powers in the weak coupling expansion can
be determined by invoking non-perturbative information from the effective
theory of three-dimensional static magnetic fluctuations.

In Ref.~\cite{Bra95}
a systematic two-step separation of the perturbatively treatable fluctuations
from the static magnetic sector has been proposed. In the first step the full
(electric {\it and } magnetic) static sector is represented by an effective
three-dimensional theory. In this theory a massive ($m_E$) adjoint scalar
field stands for the screened electric fluctuations.
In the second step this theory is matched onto an effective magnetic theory
(MQCD) with a separation cut-off $\Lambda_M$.
While the contribution of the non-static and of the static
electric modes to the free energy can be safely calculated perturbatively
(expansion parameters are $g(T)$ and $m_E/T$, respectively), the magnetic
sector is still non-perturbative and should be investigated by numerical
methods.

The high temperature free energy of the full theory can then be written in
 additive form:
\begin{equation}
f_{QCD}(T>>T_c)=f_{NS,E}(\Lambda_M)+f_M(\Lambda_M),~~~gT\geq\Lambda_M\geq g^2T.
\end{equation}
The double subscript $NS,E$ refers to the fact that the first term contains
contributions to the free energy from the non-static
as well as from static  electric type fluctuations.

The strategy of the hierarchical calculation has been tested by Braaten and
Nieto \cite{Bra95b} by reproducing from the effective electric QCD theory
the ${\cal O}(g^5)$ term in the free energy density of the SU(N) gauge theory,
which earlier has been calculated directly in the full theory by Zhai and
Kastening \cite{Zha95}. This is fully part of the term $f_{NS,E}$.

The necessary non-perturbative information to go beyond this stage comes from
the minimal MQCD theory. The purpose of our paper is to present
first results of a lattice analysis of MQCD, ie. the 3-dimensional SU(N)
gauge theory. This does provide the non-perturbative input needed to
evaluate the ${\cal O}(g^6)$
part of $f_M$, denoted in the following by $f_M^{(0)}$.

This piece of information can be derived from the simplest three-dimensional
gauge theory:
\beqn
f^{(0)}_M  = -{T\over V}\ln \left[ \int^{(M)}{\cal D}A_j^a
\exp \left(-\int d^3xL^{(0)}_{MQCD}\right)\right],
\eqn
where
\beqn 
L^{(0)}_{MQCD}  =  {1\over 4}G_{ij}^aG_{ij}^a
\eqn
is the only operator of dimension four contributing to leading order.
The gauge coupling of MQCD with the necessary accuracy is given by
$g_3^2=g^2T$. It is worth to note that the leading (${\cal O}(g^3)$)
correction to $g_3^2$ comes from the
finite wave function renormalization factor due to the integration over the
static electric field.

The determination of $f_M^{(0)}$  represents a certain interest in itself.
It provides information about the nature of quasi-particle excitations
in the high temperature phase. Similar to the situation in the electric
sector of the SU(N) gauge theories one also may expect that
at very high temperature a weakly interacting gas of
some (quasi)particles dominates the free energy density contribution of the
magnetic sector. Yet, the nature of the "constituents" of this gas is still
to be clarified.

Inspired by the
reduction strategy we have recently investigated the problem of magnetic
screening both in pure SU(2) and in SU(2)-Higgs models \cite{Kar96}.
In Landau gauge we have analyzed the vector ($A_i$) two point
correlations and determined the propagator mass from the corresponding Euclidean
propagator. The numerical results were found to be similar to those
obtained in analytical calculations from coupled gap-equations for the Higgs
and the vector channels \cite{Buc95}.
In the pure gauge theory \cite{Tep92} and in the $SU(2)$ gauge-Higgs systems
also  heavier excitations have been identified numerically
\cite{Kaj95,Per95,Phi96} and interpreted analytically \cite{Dos96}. A unified
interpretation of these rich spectra is, however, missing at present. In
particular one has to clarify whether the light excitations found so far
only in gauge dependent correlation functions really will dominate the
thermodynamics in the high temperature ideal gas limit. Understanding the
nature of (quasi)particle constituents in the high temperature limit
still is one of the major challenges in non-abelian gauge theories. The single
number $f_M^{(0)}$ offers an interesting input to this discussion as it is
sensitive to the mass of the thermodynamically relevant lightest magnetic
excitations.
\section{The 3-d magnetic free energy density}

Simple analysis of the vacuum diagrams to 4 loop order in the
3-d SU(N) gauge theory with cut-off $\Lambda_M$ leads to the following
dependence of $f_M^{(0)}$ on the magnetic separation scale:
\begin{eqnarray}
f_M^{(0)} & \hspace{-0.7em} = &
\hspace{-0.7em} T \left[ a_0\Lambda_M^3\ln{\Lambda_M\over g_3^2}+
a_1\Lambda_M^3+a_2\Lambda_M^2g_3^2+a_3\Lambda_Mg_3^4 +\left( a_4+a_4^\prime
\ln{\Lambda_M\over g_3^2} \right) g_3^6 \right] \nonumber\\
& & \hspace{-0.7em} + {\cal O}\left(g_3^8{T\over \Lambda_M^2}\right).
\label{freseries}
\end{eqnarray}
It is worth to remark, that the appearence of the gauge coupling $g_3^2$ under
the logarithms implicitly assumes the existence of an infrared scale
proportional to it.

The complete free energy density does not depend on $\Lambda_M$. Therefore the
coefficients of the terms diverging with $\Lambda_M$ should coincide with
their perturbatively calculable values. Only in this way the $\Lambda_M$
dependence can cancel against the fully perturbatively determined $f_{NS,E}$.
The finite part proportional to $a_4$, however, is fully non-perturbative,
since all loop diagrams at and beyond four loops contribute to it. This is
the quantity we would like to extract in a non-perturbative calculation on
the lattice.

Starting from ${\cal O}(g^8)$, contributions to the free energy density
are influenced also by the higher dimensional operators appearing in the
effective magnetic theory. This implies, that the ${\cal O}(g^7)$
nonperturbative contribution
reflects the ${\cal O}(g^3)$ correction of the $g_3^2=g^2T$ relation and is
still calculable within the minimal effective theory.

The actual procedure for the determination of the coefficients in
(\ref{freseries}) amounts to measuring the coefficients of the weak
coupling expansion of the internal energy density of the system.
Since the temperature dependence appears in this theory exclusively through
$g_3^2$, one finds for the energy density 
\begin{eqnarray}
\epsilon  & \hspace{-0.8em} = & \hspace{-0.8em}  {T^2\over V}{dg_3^2\over dT}
{d \over dg_3^2} \ln Z  \nonumber\\
& \hspace{-0.8em}\equiv & \hspace{-0.8em}-T^2{dg_3^2(T)\over dT}\epsilon_3.
\label{epsilona}
\end{eqnarray}
The structure of the 3-dimensional energy density and its cut-off dependence
is given by
\begin{eqnarray}
\epsilon_3 & \hspace{-0.8em} = &
\hspace{-0.8em} -a_0\Lambda_M^3g_3^{-2}+
a_2\Lambda_M^2+2a_3\Lambda_Mg_3^2+
(3a_4-a_4^\prime )g_3^4+3a_4^\prime g_3^4\ln{\Lambda_M\over g_3^2} ~.
\label{epsilon}
\end{eqnarray}

The lattice regularization of the minimal 3-d SU(N) gauge theory is defined
in the standard way:
\begin{equation}
S_{LMQCD}=\beta_3\sum_P \left(1-{1\over 2N}\Tr (U_P+U_P^\dagger )
\right),~~~\beta_3={2N\over g_3^2a},
\end{equation}
where $U_P$ denotes the Wilson plaquette variable defined in terms of
SU(N) valued variables $U_{x,i}$ \cite{wilson}.
The partition function is given by
\begin{equation}
Z_{LMQCD}=\int \prod_{x,i}{\rm d}U_{x,i}\exp (-S_{LMQCD}).
\end{equation}
The internal energy $\epsilon_3$ and the plaquette expectation value
\begin{equation}
\langle P \rangle = \left\langle 1-{1\over 2N}
\Tr (U_P+U_P^\dagger) \right\rangle=- {1\over V} {d\over d\beta_3}\ln Z
\end{equation}
can be simply related:
\begin{equation}
\epsilon_3 = -3\Lambda_M^3{\beta_3\over g_3^2}\langle P \rangle.
\label{inten3d}
\end{equation}
Here $\Lambda_M \equiv a^{-1}$ is chosen to coincide with the cut-off of
the lattice regularized theory.
\section{Plaquette Expectation Value}

The basic non-perturbative input from a lattice calculation is obtained
through an evaluation of the plaquette expectation value for a SU(N)
gauge theory. Its perturbative expansion for large $\beta$  has been
calculated up to ${\cal O}(\beta^{-2})$ for arbitrary dimensions $d$ and on finite
lattices \cite{Hel85}.\footnote{We use $\beta$ for the coupling in
arbitrary dimensions.}
\begin{eqnarray}
\langle P \rangle &\equiv& \left\langle 1-{1\over 2N}{\rm Tr}(U_P+U_P^{\dagger})
\right\rangle \nonumber \\
&=& \sum_{n} c_{n,d} \beta^{-n}  \label{pertplaq}\\
&=& (N^2 -1) I_d \beta^{-1}   +
\biggl( {(2N^2-3)(N^2-1) \over 6} I_d^2 + 4N^2(N^2-1) \alpha_d \biggr) 
\beta^{-2} \nonumber \\ 
& & +{\cal O}(\beta^{-3})~~, \nonumber
\end{eqnarray}
where $I_d = {1\over d} \bigl( 1- {1\over V} \bigr)$ and $\alpha_d$ is a
numerical coefficient, which has a weak volume dependence. On an infinite
lattice a direct evaluation gives $\alpha_4 = -0.000103$ for $d=4$
and  $\alpha_3 = -0.00095$ for $d=3$. We note that at order $\beta^{-n}$
the dominant contribution to the expansion coefficients comes from
diagrams which are proportional to $d^{-n}$. This allows to estimate
also the expansion coefficient at ${\cal O}(\beta^{-3})$ which so far has
only been evaluated in four dimensions \cite{All94}. 
These coefficients are $c_{3,4}= 0.143055$
in the case of SU(2) and 2.960467 in the case of SU(3).
Multiplying with a factor $(4/3)^3$ one finds as an estimate in three dimensions
\beqn
c_{3,3} \simeq \cases{
0.34, & SU(2) \cr
7.02, & SU(3) \cr}~~.
\label{c33est}
\eqn
The possible appearance of a logarithmic $\beta$-dependence in the 
next order expansion coefficient, $c_{4,3}$,  
follows from combining (\ref{epsilon}) and (\ref{inten3d}). Its
consequences will be discussed below.

\begin{figure}[htb]
\begin{center}
  \epsfig{file=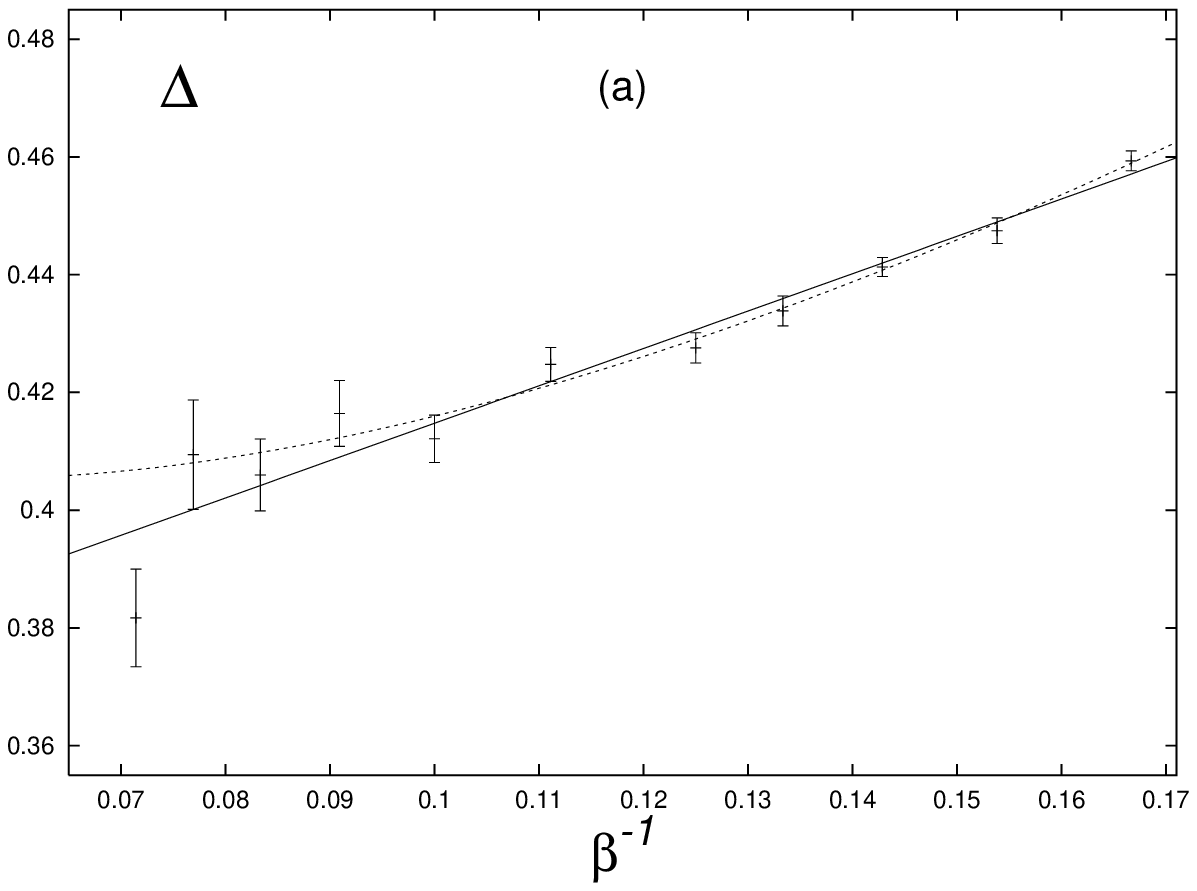, width=73mm, height=65mm}
  \hfill
  \epsfig{file=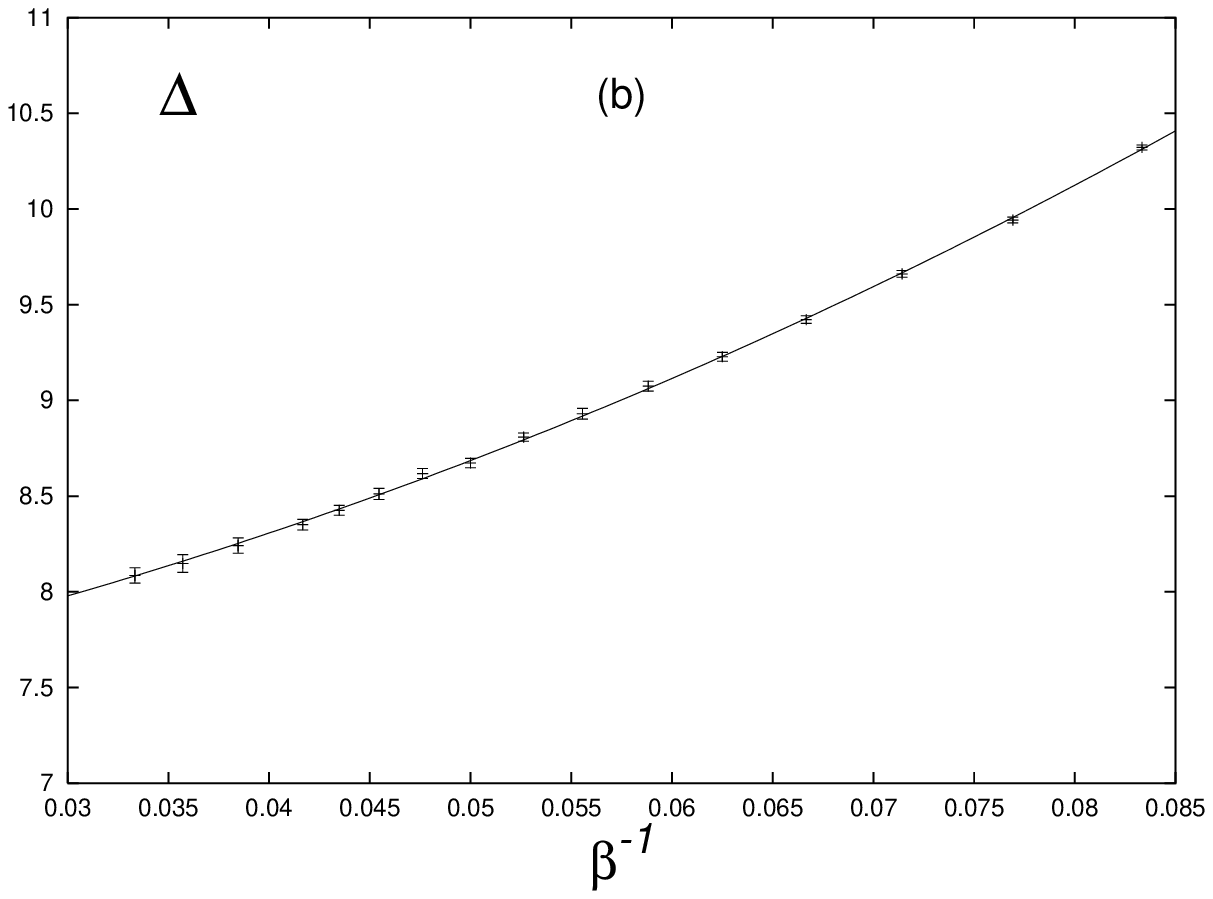, width=73mm, height=65mm}
\end{center}
\caption{The subtracted plaquette expectation values $\Delta$ as defined in
Eq.~(\ref{delta}) for the SU(2) (a) and SU(3) (b) gauge theory. The solid
curves show fits with $c'_4 =0$ (case I) and the dashed line corresponds to
$c'_4 \ne 0$ (case II). For the SU(3) case both fits coincide in the
coupling range shown.}
\label{differences.fig}
\end{figure}

As we are finally only interested in the finite part in $\epsilon_3$ our aim
is to extract the coefficient $c_{4,3}$ in the expansion of
$\langle P \rangle$.
In order to do so we subtract the known part of the perturbative expansion
from the numerical results  for the plaquette expectation values obtained from
a Monte Carlo simulation
and determine the coefficients $c_{4,3}$ and $c_{3,3}$ from a fit to these
differences. We have calculated plaquette expectation values at a large set
of $\beta_3$ values both for the SU(2) and SU(3) gauge theory, using
lattices of sizes $16^2 \times 64$ and $32^3$, respectively . Results
for $\langle P \rangle$ and the difference
\beqn
\Delta \equiv \beta_3^3 \biggl( \langle P \rangle - c_{1,3} \beta_3^{-1}
-c_{2,3} \beta_3^{-2} \biggr)
\label{delta}
\eqn
are summarized in Tables~\ref{plaqdata_su2.tab} and \ref{plaqdata_su3.tab}.
\begin{table}
\begin{center}
\begin{tabular}{|c|l|r|c|}
\hline
\multicolumn{4}{|c|}{SU(2)} \\
\hline
$\beta_3$ & \multicolumn{1}{|c|}{$\langle P \rangle$} &
\multicolumn{1}{|c|}{$\Delta$} & \# iterations \\
\hline
  6.0 & 0.1752315(78) & 0.4593(17) &  70000 \\
  6.5 & 0.1609606(79) & 0.4475(22) &  60000 \\
  7.0 & 0.1488726(47) & 0.4413(16) &  50000 \\
  7.5 & 0.1384806(60) & 0.4338(25) &  80000 \\
  8.0 & 0.1294547(50) & 0.4276(26) & 100000 \\
  9.0 & 0.1145530(39) & 0.4248(29) & 110000 \\
 10.0 & 0.1027277(40) & 0.4121(40) &  90000 \\
 11.0 & 0.0931350(42) & 0.4164(56) & 110000 \\
 12.0 & 0.0851753(35) & 0.4060(61) & 100000 \\
 13.0 & 0.0784784(42) & 0.4094(93) &  50000 \\
 14.0 & 0.0727477(30) & 0.3817(83) &  50000 \\
\hline
\end{tabular}
\end{center}
\caption{Plaquette expectation values and normalized differences as defined
in Eq.~(\ref{delta}) calculated on a $16^2 \times 64$ lattice.}
\label{plaqdata_su2.tab}
\end{table}
\begin{table}
\begin{center}
\begin{tabular}{|c|l|r|c|}
\hline
\multicolumn{4}{|c|}{SU(3)} \\
\hline
$\beta_3$ & \multicolumn{1}{|c|}{$\langle P \rangle$} &
\multicolumn{1}{|c|}{$\Delta$} & \# iterations \\
\hline
  12 & 0.2417305(77) &10.322(13) & 10000 \\
  13 & 0.2211859(68) & 9.942(15) & 10000 \\
  14 & 0.2039402(64) & 9.661(18) & 10000 \\
  15 & 0.1892309(59) & 9.423(20) & 10000 \\
  16 & 0.1765317(56) & 9.228(23) & 10000 \\
  17 & 0.1654524(52) & 9.074(26) & 10000 \\
  18 & 0.1556935(48) & 8.931(28) & 10000 \\
  19 & 0.1470324(32) & 8.808(22) & 20000 \\
  20 & 0.1392884(31) & 8.673(25) & 20000 \\
  21 & 0.1323326(28) & 8.618(26) & 20100 \\
  22 & 0.1260367(27) & 8.511(29) & 20000 \\
  23 & 0.1203172(21) & 8.426(26) & 30300 \\
  24 & 0.1150972(20) & 8.351(28) & 30500 \\
  26 & 0.1059145(23) & 8.242(40) & 20000 \\
  28 & 0.0980936(21) & 8.148(46) & 19000 \\
  30 & 0.0913523(15) & 8.085(40) & 36000 \\
\hline
\end{tabular}
\end{center}
\caption{Plaquette expectation values and normalized differences as defined
in Eq.~(\ref{delta}) calculated on a $32^3$ lattice.}
\label{plaqdata_su3.tab}
\end{table}
The difference $\Delta$ is shown in Figure~\ref{differences.fig}.
In $\Delta$ the leading powerlike divergences have been
eliminated. These would anyhow be canceled in the final physical
result for the energy density through a corresponding
perturbative calculation within the effective electric theory (NS,E).

The remaining cut-off dependence results from a possible logarithmic
cut-off dependence of the free energy density, ie. the term proportional
to $a_4^\prime$ in Eq. (\ref{epsilon}). We thus expect $\Delta$ to depend
on $\beta_3$ as follows:
\beqn
\Delta = c_{3,3} + \biggl( c_4^\prime \ln \beta_3 +
c_{4,3}\biggr) {1 \over \beta_3} + {c_{5,3} \over \beta_3^2}+ {\cal
O}(\beta_3^{-3})~~.
\label{deltamod}
\eqn
The coefficient $c_4^\prime$ is directly related to $a_4^\prime$ and
could in principle be fixed through a perturbative calculation within
the effective electric theory. We also note that $c_{3,3}$ can directly
be evaluated in lattice perturbation theory (see Eq.~\ref{c33est}).

Our numerical data for the plaquette expectation values are limited to
a $\beta$-range that varies by a factor of about $2.5$. This is not large
enough to be sensitive to the logarithmic term given above and at the
same time allow a control over the subleading corrections proportional
to $c_{5,3}$. We thus have analyzed the numerical data using two different 
fits. Since for the moment the 4-loop perturbative information on the 
coefficient of the logarithmic term as well as a direct evaluation of
$c_{3,3}$ are missing we had to fit these. In order to
see the sensitivity of $c_{4,3}$ to the variation of $c_4^\prime$ we have 
investigated also another option, corresponding to an ideal decoupling of the 
magnetic fluctuations from the rest of the system: $c_4^\prime =0$.
It will become clear in the next section that the range of $c_4$ values
obtained in this way puts already a
rather stringent bound on the mass of magnetic excitations in the plasma
phase. 
We have fitted $\Delta$ with the ans\"atze
\beqn
\Delta = \cases{
c_{3,3} + {c_{4,3} \over \beta_3} + {c_{5,3} \over \beta_3^2} &
case I~~, \cr
c_{3,3} + \biggl( c_4^\prime \ln \beta_3 +
c_{4,3}\biggr) {1 \over \beta_3} + {c_{5,3} \over \beta_3^2} &
case II~~. }
\label{deltafit}
\eqn
Results of these fits to the plaquette expectation values are summarized
in Table~\ref{tab:fits}.
\begin{table}
\begin{center}
\begin{tabular}{|c|c|c|c|c|}
\hline
  & \multicolumn{2}{|c|} {SU(2)} &  \multicolumn{2}{|c|} {SU(3)} \\
\hline
  & case I &  case II & case I &  case II \\
\hline
  $c_{3,3}$ & 0.351(5) & 0.45(5) & 7.30(11) & 8.11(20) \\
  $c_{4,3}$ & 0.635(37) & 1.4(4) & 15.2(3.6) & 99(6) \\
  $c_4'$ & -- & -0.75(35) & -- & -29.2(3.4) \\
  $c_{5,3}$ & -- & -- & 252(29) & -- \\
\hline
\end{tabular}
\end{center}
\caption{Results of the fits using the two fitting functions defined in
Eq.~(\ref{deltafit}).}
\label{tab:fits}
\end{table}
As our data for the $SU(2)$ gauge theory did not show any significant
curvature in the coupling range explored by us we have fixed $c_{5,3}$ to be
zero in both fits.
 
We note that the expansion coefficients for SU(2) gauge group are
substantially smaller than in the case of SU(3). This is in agreement
with the expected $N$-dependence for the expansion coefficients for
different $SU(N)$ groups. In fact, the fitted coefficient $c_{3,3}$
agrees well with the estimate given in Eq.~(\ref{c33est}). This estimate
does seem to favour the fit I.

While the coefficient $c_{4,3}$ only changes by a factor 2 in the case of
$SU(2)$ for both fits, the variation is about twice as large in the case of
$SU(3)$.  However, it will become
clear from the discussion in the following section that already our
present estimates for $c_{4,3}$ are rather restrictive for the effective
mass of magnetic excitations in the plasma phase, since this latter is only
sensitive to its cubic root.

\section{The ${\cal O}(g^6)$ coefficient and its interpretation}

Combining (\ref{inten3d}),(\ref{pertplaq}),(\ref{delta}) and 
(\ref{deltamod}) for case II  we relate the lattice and continuum coefficients
in the following way:
\begin{eqnarray}
a_4^\prime & = &-{1\over (2N)^3}c_4^\prime ,\nonumber\\
a_4 & = &-{1\over (2N)^3}\left(c_{4,3}+c_4^\prime (\log (2N)+{1\over 3})\right).
\end{eqnarray}
Case I corresponds to setting $c_4^\prime =0$. The values of $c_{4,3}$ and 
$c_4^\prime$ provide the input from the magnetic sector for the determination
of the ${\cal O}(g^6,g^6\log g)$ terms in the weak coupling series of the
free energy density.

The numerical values  of the previous section lead to:
\begin{equation}
a_4^{(I)} =\cases{
-(0.010 \pm 0.001), & SU(2)\cr
-(0.07  \pm 0.02),  & SU(3)\cr}\nonumber\\
\end{equation}
and
\begin{equation}
a_4^{(II)}=\cases{
-(0.002 \pm 0.016), & SU(2)\cr
-(0.17\pm 0.06), & SU(3) \cr}~~.
\end{equation}
For the SU(2) group the estimated value for $a_4$ in case II is
compatible with zero within errors. Clearly,
this means that the accuracy of our simulation in the SU(2) case is yet 
insufficient to become sensitive to logarithmic corrections. We note,
however, that in the SU(2) as well as in the SU(3) case the inclusion of 
a possible logarithmic term in the fit at most increases the value of $a_4$
by a factor of about 2.5. 

Let us try to get some feeling for the magnitude of $a_{4}$ and its
interpretation in terms of physical excitations.
The 3-d theory represents the magnetic fluctuations of the high-T
(3+1)-dimensional gauge theory.
It is natural to expect that the $O(g^6)$ contribution to the free energy
can be viewed also as resulting from a weakly interacting
(almost ideal) gas of some pseudo-particles of mass $m_M\sim g^2 T$
and degeneracy $N_D$.
This is quite analogous to the electric sector, where the $O(g^3)$
contribution to the free energy density is just the contribution of a massive
free gas of pseudo-particles with mass $m_E\sim gT$.
The well-known 3-d, 1-loop vacuum energy has the cut-off independent part:
\begin{equation}
\epsilon_3=-{3N_Dm_M^2\over 12\pi}{dm_M\over dg_3^2}=-{N_Dm_M^3\over 4\pi
g_3^2}.
\label{epsilon3}
\end{equation}
In the last equality the proportionality of $m_M$ to
$g^2_3$ has been exploited. 

The picture of an ideal gas of pseudo-particles cannot account for the 
logarithmic piece. If present at all this term would arise from higher order 
interactions of the effective degrees of freedom. 
The natural combination appearing in the argument
of the logarithm is the ratio of the ultraviolet scale (the lattice constant)
and a dynamically generated infrared scale $m_G^{-1}$. This scale is 
proportional to $g_3^2$: $m_G=c_Gg_3^2$. This fixes the separation of the
terms proportional to $g^6$ and to $g^6\log g$, necessary for the 
quantitative investigation of the non-logarithmic piece:
\begin{equation}
a_4^{nonpert}=-{1\over (2N)^3}\biggl[ c_4+c_4^\prime ({1\over 3}+\ln (2Nc_G))
\biggr].
\label{a4nonpert}
\end{equation}
The comparison of (\ref{epsilon3}) to (\ref{inten3d}) via (\ref{pertplaq}) 
leads to
\begin{equation}
m_M=\left(- {12\pi a_4^{nonpert}\over N_D}\right)^{1/3}g_3^2.
\label{mass}
\end{equation}

At least two simple cases can be put forward for the pseudo-particle
excitations resulting from the high-T magnetic modes. With the ansatz given
by Eq.~(\ref{mass}) they correspond to different choices for the degeneracy
factor $N_D$. The extreme cases clearly are to set $N_D$ equal to the number
of gluonic degrees of freedom or to unity. The first case may be
considered as being the analogous interpretation to the electric sector,
while the second assumes that there do exist only color singlet magnetic
modes above $T_c$:

\noindent
i) Adjoint SU(N) multiplet of screened gluons

Our previous determination of the magnetic screening mass of the
gluons has been performed in Landau gauge \cite{Kar96}. For reasons of
comparability we continue to use this gauge. Then the magnetic mass and the 
infrared regularization mass  are naturally identified. The number of degrees 
of freedom in this gauge is then given by $N_D=2(N^2-1)$. The solution $c_G$ 
of the equation (\ref{mass}) leads to the following mass estimates:
\begin{equation}
m_{gluon}^{(I)}= \cases{
(0.397 \pm 0.008)~g_3^2, & SU(2)\cr
(0.55 \pm 0.04)~g_3^2, & SU(3)\cr}
\end{equation}
and
\begin{equation}
m_{gluon}^{(II)}=(0.78\pm 0.29)g_3^2,~~~~ {\rm SU(3)}.
\end{equation}
This later value is still compatible within error bars with case I. 
Within the errors of the simulation the SU(2) result is perfectly compatible
with the gluon mass calculations presented in Ref.~\cite{Kar96}. We also note 
that the ratio of the SU(2) and SU(3) mass values for case I is very close to 2/3,
which is expected from the 1-loop gap equation approach of \cite{Buc95},
since the 1-loop diagrams contribute a quantity proportional to $N$
to the self energy.

\vfill\eject
\noindent
ii) Scalar, SU(N) singlet glueballs ($N_D$=1)

One arrives at the following prediction for the glueball mass:
\begin{equation}
m_{singlet}^{(I)}= \cases{
(0.721 \pm 0.014)~g_3^2, & SU(2)\cr
(1.38  \pm 0.11)~g_3^2,  & SU(3)\cr}
\end{equation}
and
\begin{equation}
m_{singlet}^{(II)}=(2.08\pm 0.67)g_3^2,~~~~ SU(3).
\end{equation}
For the determination of $a_4$ in case II one requires as additional input
the infrared regularization scale. For the SU(3) case we used the
measured magnetic screening mass, determined for SU(2)
\cite{Kar96}, and scaled it up by a factor 3/2, that is we used $c_G\sim 0.6$
in Eq.~(\ref{a4nonpert}).

These values can be partly compared with existing numerical results on 3-d
glueballs. For the SU(2) theory the lowest glueball mass \cite{Tep92} 
calculated in numerical simulations $m_{glueball}= 6.34(6)g_3^2$ is clearly 
much larger than our estimate. Additional (higher) glueball
states would make this discrepancy even more dramatic. Also the estimate
of zero temperature glueball masses in four dimensions and estimates
of finite temperature glueball screening masses in the (3+1)-d SU(3) gauge
theory lead to larger values \cite{Gro94}, though the value found in case
II within errors is on the edge of being compatible with the MC estimate.

The above comparisons seem to present a rather strong evidence
against an interpretation of the thermodynamics of magnetic fluctuations 
in terms of singlet
excitations with a mass similar to a typical zero temperature glueball mass.
This analysis suggests that the relevant thermodynamic
degrees of freedom at high temperatures in the magnetic sector of the
non-Abelian plasma are screened elementary gluons.

\section{Conclusions}

We have performed a first analysis of the non-perturbative contribution
to the ${\cal O} (g^6)$ coefficient of the free energy of a finite
temperature SU(N) gauge theory. 
We stress that the analysis we have presented here is not complete. 
At the present stage we tried to investigate whether a lattice
calculation can provide the necessary non-perturbative
information that is needed to determine the
complete $O(g^6)$ term in the free energy of an $SU(N)$ gauge theory. Our
calculation shows that this is indeed feasible. 
Combined with a rather straightforward calculation of the expansion
coefficient $c_{3,3}$ within lattice perturbation theory and
an analytic determination of $c_4^\prime$ a much better determination of
$a_4$ can be achieved already on the basis of the numerical results presented 
here. The quality of 
the numerical input may also be further improved by
exploring improved actions which reduce the size of the strongly cut-off 
dependent parts of
observables like the expectation value of the Euclidean action. 
It is very encouraging, however, that already at the present stage we can 
distinguish between qualitatively different ideas on the nature of the
high temperature magnetic excitations.

\medskip
\noindent
{\bf Acknowledgements:}
The computations have been performed on Connection Machines at the
H\"ochstleistungs\-rechenzentrum (HLRZ) in J\"ulich, the University of
Wuppertal and the Edinburgh Parallel Computing Center (EPCC).
We thank the staff of these institutes for their support.
The work of FK has been supported through the Deutsche
Forschungsgemeinschaft under grant Pe 340/3-3. JR has partly been supported
through the TRACS program at the EPCC. A.P. acknowledges a grant from OTKA.

\end{document}